\documentclass[conference]{IEEEtran}
\IEEEoverridecommandlockouts
\usepackage{cite}
\usepackage{amsmath,amssymb,amsfonts}
\usepackage{algorithmic}
\usepackage{booktabs}
\usepackage{graphicx}
\graphicspath{ {./images/} }

\usepackage{float}
\usepackage{xurl}
\usepackage{textcomp}
\usepackage{xcolor}
\usepackage{multirow}
\usepackage{tikz}
\usepackage{pgfplots}
\pgfplotsset{compat=1.17}
\usepackage{filecontents}
\def\BibTeX{{\rm B\kern-.05em{\sc i\kern-.025em b}\kern-.08em
    T\kern-.1667em\lower.7ex\hbox{E}\kern-.125emX}}
\begin{document}

\title{Towards Real-Time Interpolation for Enhanced AUV Deep Sea Mapping\\}

\author{\IEEEauthorblockN{Devanshu Saxena}
\IEEEauthorblockA{\textit{Department of Computer Science} \\
\textit{William \& Mary}\\
Williamsburg, VA, USA \\
dsaxena@wm.edu}
}

\maketitle

\begin{abstract}
Approximately seventy-one percent of the Earth is covered in water. Of that area, ninety-five percent of the ocean has never been explored or mapped. There are several engineering challenges that have prevented the exploration of the deep ocean through human or autonomous means. These challenges include but are not limited to high pressure, cold temperatures, little natural light, corrosion of materials, and communication. Ongoing research has been focused on trying to find optimal and low-cost solutions to effective communication between autonomous underwater vehicles (AUVs), and the surface or air. In this paper, an architecture is introduced that utilizes an edge computing approach to establish computation nearer to the source of data, allowing further exploration of the deep ocean. Taking the most common interpolation techniques used today in the field of bathymetry, the data are tested and analyzed to find the feasibility of switching from CPU to GPU computation. Specifically, the focus is on writing efficient interpolation algorithms that can be run on low-level GPUs, which can be carried onboard AUVs as payload. The code can be found at: \url{https://github.com/devsaxena974/AUV-Real-Time-Interpolation}
\end{abstract}


\section{Introduction}
Exploration of the deep ocean is crucial to study climate change, energy, ocean conservation, and human health. Further knowledge of what lies in the dark depths of the ocean seabed could help discover renewable energy sources, advance natural medicine research, and protect the ocean ecosystem\cite{b4}.

Before introducing the architecture, there are several constraints that have shaped previous and current implementations of solutions to deep ocean exploration. Many of the underlying ones stem from physical and chemical phenomena, but also the computational expenses of processing data. Although this paper does not focus on the physical design aspects of the devices used for exploration, these challenges can help understand why a real time interpolation system would be useful. The following subsections provide motivation for the research in this study.

\subsection{Temperature and Pressure}

In ocean depths past six thousand meters, the pressure of water reaches approximately 596 atm. This is problematic for most electronics that need to be mounted on the AUV such as cameras, lights, and computers. Since most are built to be used above the water, they operate in 1 atm of air. To combat this, the most practical solution is to enclose the technological devices in an air-filled housing that will not collapse under such intense pressure.

The temperature at such depths in the ocean can reach to around four degrees Celsius. This also poses a problem for dissimilar materials - materials that are difficult to join as a result of their chemical or physical compositions - because they may shrink or expand at various rates when the temperature falls\cite{b2}.

\subsection{Corrosion and Darkness}

In terms of what materials to use, saltwater is an electric conductor and will accelerate the corrosion of metals that are immersed in it. In their development, AUVs and remotely operated vehicles (ROVs) are engineered with materials that are low on the Galvanic scale.

Additionally, there is very little natural light below 200 meters in the ocean and no sunlight below 1000 meters. With underwater creatures, terrain, and other organisms to consider, AUVs use the help of sonar to determine position, listen for the seabed below, and record mapping data.\cite{b2}.

\subsection{Sonar Mapping and Communication}

AUVs use multi-beam echosounders \cite{b9} that are mounted to the bottom of the ship to collect mapping data. They emit sound into the water, which hits the seafloor and echoes back to the receiver. The time and frequency are returned, which can be used to calculate the depth at that point. Once this data is processed, with current architectures, it is either stored onboard or sent back to a mother-ship via EM waves or some sort of tether.

Traditionally, some form of radar or EM waves would be used for autonomous vehicles but radio waves do not travel very far through water and are very susceptible to noise. This leaves solutions that are separated into categories that involve a tethered connection and untethered connections \cite{b2}.

Tethered connections to vehicles involve cable(s) that are coaxial, in twisted pairs, or fiber-optic. Since this is a direct connection to a mother-ship on the surface, tethered solutions have a high data transfer rate but may have lower ranges. Outside of these considerations, a long cable that extends into the ocean could face entanglement, breaking, fraying, and encounters with creatures.

Untethered connections involve communication from the vehicle to another device via radio waves, acoustic waves, or optical methods such as lasers or LEDs. Having already discussed the drawbacks of radio wave communication, acoustics are widely used because of their long range abilities. However, most acoustic solutions succumb to high noise and low bandwidth in data transfer. The relatively newer idea of optical communications, with some research in space exploration being pioneered by NASA, solves this issue by providing higher bandwidth although within closer distances\cite{b1}. Table I and Table II show the comparisons between the communication types\cite{b1}.

\begin{table}[h!]
\centering
    \begin{tabular}{c|c|c|c}
        
        Type & Technology & Data Rate & Range \\ \hline
        Tether &  Twisted Cables & 10 Gb/s & 100m \\
        Tether &  Optical Fiber & 10 Tb/s & 10,000m \\
        Wireless & Acoustic & 10 Kb/s & 1000m\\
        Wireless & Optical & 100 Mb/s & 100m \\
        Wireless & Radio & 10 Mb/s & 1m
    \end{tabular}
\vspace{0.2cm}
\caption{Communication technologies, their types, their maximum data rates, and their maximum ranges}
\label{table:1}
\end{table}
\vspace{0.2cm}

\begin{table}[h!]
\centering
    \begin{tabular}{c|c|c}
        
        Tech & Use & Availability \\ \hline
        Twisted Cables & control/nav/video & commercial \\
        Optical Fiber & control/nav/video & commercial \\
        Acoustic & control/nav & commercial\\
        Optical & control/nav/video & research \\
        Radio & control/nav & commercial
    \end{tabular}
\vspace{0.2cm}
\caption{Communication technologies, their uses, and their availability}
\label{table:1}
\end{table}

\subsection{Proposed Solution}

Keeping all these constraints in mind, a new scheme that minimizes the number of data that is being transferred would be incredibly beneficial. The goal is to process data efficiently nearer to the device in the deep ocean and minimize transfer up to the surface or shore. The AUV could then use fiber optic cables, lasers, and acoustic modems to communicate all the way from the deep ocean to a mother ship on the surface.

The paper will go into the specifics of the new interpolation algorithms built to be run on an onboard GPU in real-time. This will keep the AUV underwater longer, enhance collected data, and minimize the processing that takes place outside of the AUV. Processed data can then be transferred in many different ways and using many different device architectures. One such example of how data can be transferred away from the AUV is explained briefly in the following paragraph. However, note that the focus of this paper is not on this architecture, but rather on building the interpolation algorithms that perform computation on the AUV.

\begin{figure}
    \centering
    \includegraphics[width=0.8\linewidth]{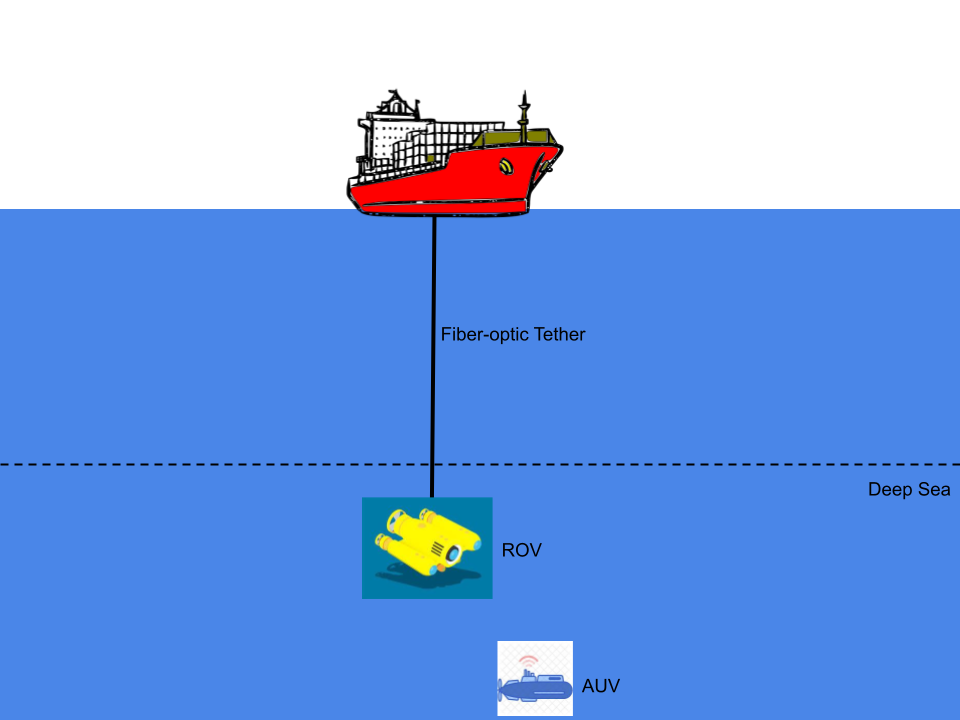}
    \caption{Example of data transfer architecture upon implementation of real time interpolation on an AUV}
    \label{fig:newArchitecture}
\end{figure}

On the surface, a mother ship stands to receive the gathered data and send any necessary, minimal instructions to an ROV. On the second level, approximately 6000 meters below sea level, this intermediary ROV acts as a translator between the mother ship and the AUV nearer to the seabed. The ROV will be tethered to the mother ship. Finally, on the third and deepest level, the untethered AUV collects data from the deep ocean, processes it in real time, and sends it to the intermediary ROV through wireless methods. A diagram of what this may look like is shown in Fig. \ref{fig:newArchitecture}

\section{Related Work}
Before taking a look at the implementation of the real time interpolation, it is worth talking about the most recent advancements in the field to adapt the study accordingly.

\subsection{GFOE Deep Discover}
In 2022 the Global Foundation for Ocean Exploration had built one of the most capable ROVs to date called the Deep Discover (D2). The D2 boasted some great specifications with an ability to drop down to 6000 meters below sea level, used 27 LED lights, high definition cameras, a suction sampler, a manipulator arm, coral cutters, and a temperature probe. Its purpose was to collect videos, physical samples, and additional oceanic data for use by scientists.

In terms of the architecture, the D2 was tethered to a camera sled, Seirios. Seirios lit up the area under D2 and allowed D2 to safely crawl the seabed. Seirios was then tethered to a NOAA ship, Okeanos Explorer, with a six-mile-long cable made of steel. This model allowed pilots of the D2 to safely guide the ROV while it explored even the darkest parts of the deep ocean. The data transfer pipeline was through the tethered connections and allowed only the most crucial data to be transferred through the tethers without the D2 needing to surface. \cite{b2}

\subsection{Opensea IQ Edge}

A private company called Greensea Systems made waves in the computing aspect of deep ocean exploration by using untethered, commercially available, ROVs loaded with batteries, an acoustic modem, and revolutionary onboard software. They managed to put a parallelized NVIDIA edge platform on the ROV which allowed the ROV to process sonar and video data while also maintaining its navigation and communication capabilities.

Similar to the D2 from GFOE, Greensea's ROV only sent the most crucial information to human operators and did most of the data processing onboard. This method allowed the use of acoustic modems, as the amount and frequency of data were drastically reduced. The company ran a successful test of their system in March 2023. \cite{b3}

Both of these advances in current technology prove that real-time interpolation onboard ann AUV is possible and would be extremely beneficial.

\section{Interpolation Methods}

Now equipped with GPU compute capability, AUVs have the potential to process and interpolate bathymetry data faster. The following subsections describe the specific design choices made to implement the interpolation functions on the GPU and compare them against their CPU counterparts. For all of these methods, a given cell structure of existing points is assumed, and then an interpolated point is computed based on that cell. Doing this computation for multiple cells, the aim is to generate thousands of these points inside a larger grid. This is represented visually in Fig. 1.

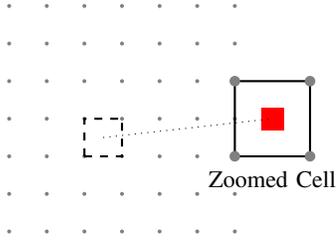
\begin{figure}
\centering
\begin{tikzpicture}[font=\small]

  % --- Mini-Grid: Smaller version of the overall grid ---
  % We draw a grid of gray intersection points over a larger area,
  % using a reduced scale so it appears in miniature.
  \begin{scope}[scale=0.5]
      \def\xmin{0}
      \def\xmax{6}
      \def\ymin{0}
      \def\ymax{6}
      
      % Draw grid intersection points as small gray dots
      \foreach \x in {\xmin,...,\xmax} {
         \foreach \y in {\ymin,...,\ymax} {
            \fill[gray] (\x,\y) circle (1.5pt);
         }
      }
      
      % Draw a dashed outline around the example cell (cell from (2,2) to (3,3))
      \draw[dashed, thick] (2,2) rectangle (3,3);

  \end{scope}

  % --- Zoomed View: Detailed drawing of the highlighted example cell ---
  % This part shows the cell in full size and includes a red square in the center.
  \begin{scope}[shift={(3,1)}] % Shift to position the zoomed cell separate from the mini-grid
      % Draw a full-scale cell outline (1x1) for the zoomed cell
      \draw[thick] (0,0) rectangle (1,1);
      
      % Draw the grid intersection points (cell corners) in the zoomed cell
      \fill[gray] (0,0) circle (2pt);
      \fill[gray] (1,0) circle (2pt);
      \fill[gray] (0,1) circle (2pt);
      \fill[gray] (1,1) circle (2pt);
      
      % Place a red square to indicate the interpolated point.
      % Here the red square is a 0.3 × 0.3 square centered in the cell.
      \fill[red] (0.35,0.35) rectangle (0.65,0.65);
      
      % Label for clarity
      \node at (0.5,-0.3) {Zoomed Cell};
  \end{scope}

  % --- Connect the Views ---
  % In the mini-grid (with scale=0.5) the highlighted cell runs from (2,2) to (3,3),
  % so its center is at (1.25,1.25). In the zoomed view (shifted by (3,1)),
  % the center of the cell is at (0.5,0.5) relative to that shift, or absolute (3.5,1.5).
  \draw[dotted] (1.25,1.25) -- (3.5,1.5);

\end{tikzpicture}
\caption{Zoomed-in view of a single interpolated value inside a larger grid. The red square represents the interpolated value}
\end{figure}

For this application, three different interpolation methods commonly used for bathymetric data are tested: Bilinear, Cubic Spline, and kriging. There are many variants and forms of each method that can be easily implemented on top of the base functionality implemented in the codebase for this paper. Each method is introduced briefly below with the specifications used in this research and sources for further reading.

\subsection{Bilinear Interpolation}

The bilinear interpolation scheme combines two linear interpolations, one along the longitude coordinates, and one along the latitude coordinates, to compute an elevation estimate in a 2-dimensional grid. Assuming an individual cell in the collected data that has 4 pre-existing points and elevation values given by $Q_{ij}$, the following notation is used: 
\begin{align*}
    &(x_1, y_1), f(x_1, y_1) = Q_{11}\\
    &(x_2, y_1), f(x_2, y_1) = Q_{21}\\
    &(x_1, y_2), f(x_1, y_2) = Q_{12}\\
    &(x_2, y_2), f(x_2, y_2) = Q_{22}
\end{align*}

In order to estimate elevation at a new $(x, y)$, the following base formula for bilinear interpolation \cite{b10} is used:

\begin{equation*}
    \begin{split}
        f(x, y) &= \frac{(x_2 - x)(y_2-y)}{(x_2-x_1)(y_2-y_1)}Q_{11} + \frac{(x - x_1)(y_2-y)}{(x_2-x_1)(y_2-y_1)}Q_{21}\\
        &+ \frac{(x_2 - x)(y-y_1)}{(x_2-x_1)(y_2-y_1)}Q_{12} + \frac{(x - x_1)(y-y_1)}{(x_2-x_1)(y_2-y_1)}Q_{22}
    \end{split}
\end{equation*}

This method will produce a weighted average of the 4 surrounding points. The closer that a point is to the target, the greater influence it will have on the interpolated result. Doing this repeatedly for several cells inside a larger "grid", required points can be constructed inside any given grid.

\subsection{Cubic Spline Interpolation (Catmull-Rom)}

The Catmull-Rom cubic spline interpolation \cite{b5} works similarly but uses 16 neighboring grid points. The method first interpolates in x on each of 4 rows, and then in y on each of 4 columns using the following cubic interpolation formula, using the Catmull-Rom variant:

\begin{equation*}
    \begin{split}
        C(t) &= \frac{1}{2}(2P_1 + (-P_0 + P_2)t + (2P_0 - 5P_1 + 4P_2 - P_3)t^2\\
        &+ (-P_0 + 3P_1 - 3P_2 + P_3)t^3)
    \end{split}
\end{equation*}

With more input data, the cubic spline interpolation scheme is expected to perform with better accuracy but also needs to query a greater number of points, increasing memory access times.

In order to simulate a real world application, the cubic spline implementation in this study falls back to a bilinear calculation if enough valid neighbors - candidate points that are not NaN values - are found.

\subsection{Ordinary Kriging}

The final interpolation scheme is the most accurate, and also the most commonly used for bathymetric applications due to its sophisticated weighting ability. Since this is likely the most foreign method of the three to readers, the method's construction is explained in greater depth than the previous ones.

There are 2 main principles that govern kriging algorithms:

\begin{enumerate}
    \item Datum close in geological distance to the unknown should get larger weight
    \item Data close together are redundant and should share their weight
\end{enumerate}

To quantify these principles, a variogram is used \cite{b6} to model the correlation in distance and elevation between any 2 arbitrary points. Let $x$ be the geographic latitude and longitude of the point and let $y$ be the elevation at that point. Then, the variogram function is given by

\begin{equation*}
    \gamma(h) = C_0(1-exp(-\frac{h}{a})
\end{equation*}

where $C_0$ represents the sill, $a$ represents the range, and $h$ is the separation distance. The rest of the kriging algorithm \cite{b7} can then be expressed as

\begin{equation*}
    \gamma(x_i, x_j) \Lambda = \gamma(x_o, x_i)
\end{equation*}

where $\Lambda$ is the matrix of weights pertaining to each neighboring point. For multiple points, such as the grids used in this study, the following matrix calculation can be solved using Gaussian elimination. Here is what that may look like for a grid consisting of 3 points (the implementation in this study uses a 5x5 grid):

\begin{equation*}
    \begin{split}
        \begin{pmatrix}
        \sigma_{11} & \sigma_{12} & \sigma_{13} \\
        \sigma_{21} & \sigma_{22} & \sigma_{23} \\
        \sigma_{31} & \sigma_{32} & \sigma_{33}
        \end{pmatrix}
        \begin{pmatrix}
        \lambda_{11} & \lambda_{12} & \lambda_{13} \\
        \lambda_{21} & \lambda_{22} & \lambda_{23} \\
        \lambda_{31} & \lambda_{32} & \lambda_{33}
        \end{pmatrix}
        =
        \begin{pmatrix}
        \sigma_{10} & \sigma_{10} & \sigma_{10} \\
        \sigma_{20} & \sigma_{20} & \sigma_{20} \\
        \sigma_{30} & \sigma_{30} & \sigma_{30}
        \end{pmatrix}
    \end{split}
\end{equation*}

Having solved for the weights, the estimated elevation is given by

\begin{equation*}
    \hat{z}(x_0) = \hat{\lambda_1}z(x_1) + \hat{\lambda_2}z(x_2)+...+\hat{\lambda_n}z(x_n)
\end{equation*}

Similarly to cubic-spline interpolation, the kriging method also falls back to bilinear interpolation if enough valid candidate neighbors are not found.

\subsection{Additional Considerations}

The choice to use basic implementations of each interpolation method was a deliberate one, as the performance and accuracy of each one is tested based on original functionality. Although fancier variants and changes can be made to provide better metrics, only the bare-bones implementations are considered in this study.

As seen later in the testing results, each interpolated point is dependent on the neighbors around it. Often in real-world applications, neighbors may not be valid or may be incorrect. To avoid this, two approaches have been taken to mitigate errors in interpolation calculations:

\begin{enumerate}
    \item A helper method is used to find the nearest $n$ valid candidate neighbors for each method, where $n$ depends on the number of neighbors required by each of the three interpolation methods
    \item As stated earlier, a fallback is implemented in the cubic spline and kriging implementations if no valid candidate neighbors are found
\end{enumerate}

Although this functionality mitigates most cases where an interpolated value can not be found due to its neighbors, there are some edge cases where methods still can not calculate a value. This is most common in the bilinear interpolation method, as there is no fallback for its implementation. For specifics on how each of these methods are implemented, readers should refer to the codebase \cite{b8}.

\section{Computational Design}

With mathematical methods laid out, the computational part of this study is discussed. The structural design comprising each point and grid is universal across the CPU and GPU implementations, coded as objects and structs that can be used on both machines. Each of the interpolation methods then has a CPU implementation in C++ and a GPU implementation in C++/CUDA.

A main control file will read in an existing grid of data obtained. Then, the CPU and GPU functions are called and timed to measure the efficiency and accuracy of the results \cite{b8}. Some testing results will be discussed in section V. In a real-time environment, the interpolation will be performed solely on the GPU, and only control functions will require the CPU. This design choice will minimize sequential processing while preserving as much memory as possible. Although there are many different ways to implement the interpolation methods on the GPU, the code in this project takes advantage of memory coalescing but assumes no shared memory exists on the GPU due to hardware limitations.

\section{Experimental Findings}

This leads to the results of implementing the aforementioned interpolation methods on the CPU and GPU, the testing design, and the evaluation metrics.

\subsection{Device Specifications}

Before seeing the results and talking about the testing setup, it is important to note that performance metrics depend heavily on the GPU device's capabilities and will have different outcomes should the device change. The GPU device used for these tests has specifications given in Fig. \ref{fig:GPUSpecs}, and the CPU has specifications given in Fig. \ref{fig:CPUSpecs}:\\

\begin{figure}
    \centering
    \begin{tabular}{ |p{3.5cm}|p{3cm}|  }
     \hline
     \multicolumn{2}{|c|}{12th Gen Intel Core i5} \\
    
     \hline
     Total RAM   & 16 Gb\\
     \hline
     Cores&   10 Cores\\
     \hline
     Total Threads & 12\\
     \hline
    \end{tabular}
    \caption{Specifications of the CPU device used in this study}
    \label{fig:CPUSpecs}
\end{figure}

\begin{figure}
    \centering
    \begin{tabular}{ |p{3.5cm}|p{3cm}|  }
     \hline
     \multicolumn{2}{|c|}{NVIDIA GeForce MX550} \\
    
     \hline
     Total Global Mem   & 2048 Mb\\
     \hline
     Cores&   1024 CUDA Cores\\
     \hline
     Max Threads per Multiprocessor & 1024\\
     \hline
     Max Dim Size of thread block   & (1024, 1024, 64)\\
     \hline
    \end{tabular}
    \caption{Specifications of the GPU device used in this study}
    \label{fig:GPUSpecs}
\end{figure}

\subsection{Testing Design}

For this study, the layout of the original data grid is very important in demonstrating how well the interpolation schemes perform on both the CPU and the GPU. Consider a grid where each latitude is fully populated with elevation values for each corresponding longitude. The purpose of interpolating on such a grid, referenced in the future as Grid A, is to fill in entire missing latitudes between any two successive existing ones. 

This would be useful in real-time applications where data frequency can be artificially increased due to hardware or geographical limitations. The original grids for Grid A tests are generated artificially and fed into the interpolation schemes.

The second case, and perhaps the most applicable of the two on AUVs, considers an original grid with missing longitude values randomly distributed throughout. This case tests filling in gaps that replicate where any noise or hardware limitations may have prevented accurate data collection. The tests for this case are referenced as Grid B.

As Grid B cases are more realistic, this study uses recorded mapping data obtained from the General Bathymetric Chart of the Oceans (GEBCO) \cite{b11} as the original grid. Data on the following regions were collected for Grid B testing:

\begin{enumerate}
    \item Mid-Atlantic Ridge: A ridge in the North Atlantic separating the American and Eurasian tectonic plates\cite{b12}
    \item East Pacific Rise: A similar, mid-ocean rise, at a divergent tectonic plate separating the Pacific plate from the North American Plate\cite{b12}
    \item Mariana Trench: Crescent-shaped trench located in the western Pacific\cite{b13}
    \item Kerguelen Plateau: A large igneous province on the Antarctic plate\cite{b14}
\end{enumerate}

Each of these regions vary greatly in their seafloor elevations as they are either tectonically active zones, or oceanic plateaus. Being highly variant, they serve as great test cases for interpolation methods, and are used as original grids in Grid B tests. Once loaded in, a fraction of data is randomly removed to simulate a case in which data needs to be interpolated and filled in. The following subsections discuss the results of Grid A and Grid B tests performed in this study.

\subsection{Grid A Testing}

The Grid A tests can be evaluated by looking at the visual bathymetry graphs produced before and after interpolation, as well as the runtime for the CPU and GPU interpolation methods on various batch sizes.

\begin{figure}[htbp]
  \centering
    \includegraphics[width=\linewidth]{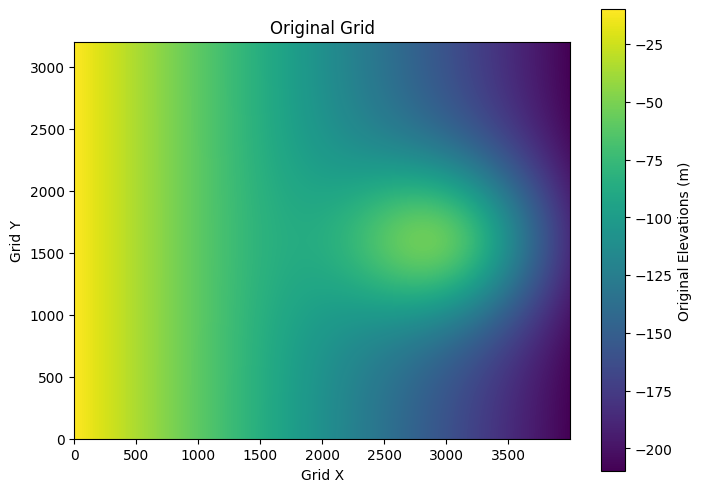}
  \caption{Original grid elevation map for Grid A testing.}
  \label{fig:gridAOriginal}
\end{figure}

\begin{figure}[htbp]
  \centering
    \includegraphics[width=\linewidth]{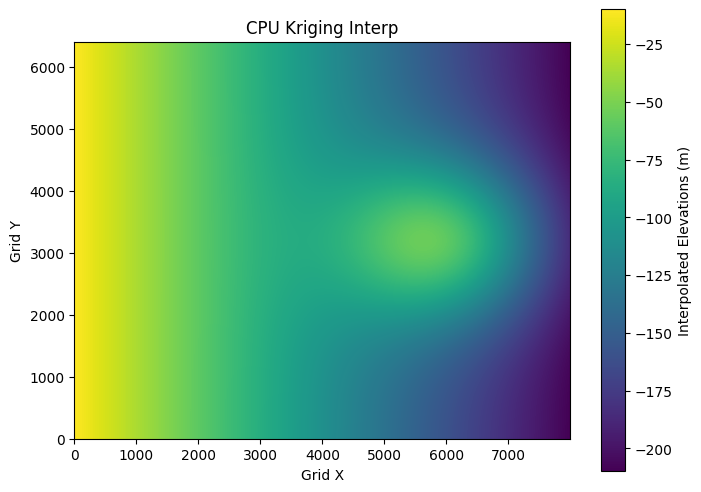}
    
    \includegraphics[width=\linewidth]{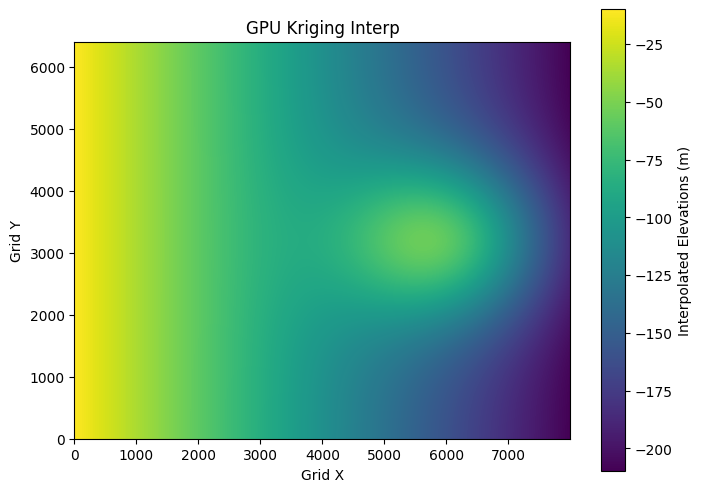}
  \caption{CPU (top) and GPU (bottom) ordinary kriging elevation maps for Grid A testing.}
  \label{fig:gridAInterpolated}
\end{figure}

Since the testing was aimed at increasing the data frequency in between preexisting rows, there is not much visual difference in the elevation maps before and after interpolation. That is a good thing though, because this means the interpolation methods are adhering to the neighboring data rather than making errors and computing elevations that vary from the surrounding points. The original grid is shown in Fig. \ref{fig:gridAOriginal} and the interpolated grids are shown in Fig. \ref{fig:gridAInterpolated}

\begin{figure}[!t]
  \centering
  \begin{tikzpicture}
    \begin{axis}[
      width=\columnwidth,
      height=0.6\columnwidth,
      xlabel={\# Interpolation Points},
      ylabel={Time (ms)},
      xmode=log,
      grid=both, grid style={dashed,gray!30},
      legend style={at={(0.02,0.98)},anchor=north west,font=\footnotesize},
      cycle list name=exotic, mark options={solid}
    ]
      \addplot table[x=points,y=cpu_bilin,col sep=comma] {data/grid_A_runtimes_averaged.csv};
        \addlegendentry{CPU}
      \addplot table[x=points,y=gpu_bilin,col sep=comma] {data/grid_A_runtimes_averaged.csv};
        \addlegendentry{GPU}
    \end{axis}
  \end{tikzpicture}
  \caption{Semi-log plot of bilinear interpolation for Grid A testing}
  \label{fig:bilinear_perf}
\end{figure}
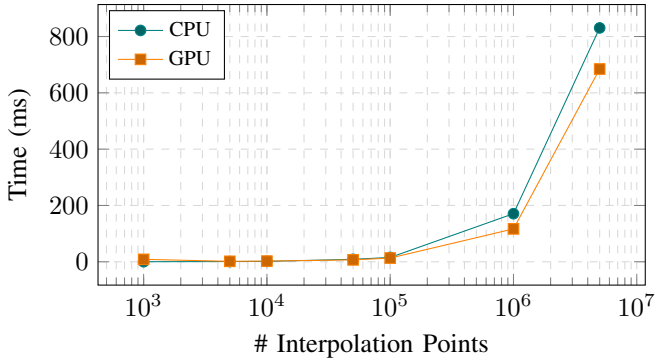

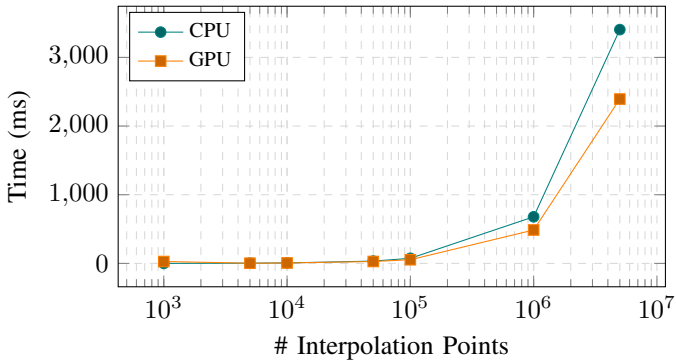
\begin{figure}[!t]
  \centering
  \begin{tikzpicture}
    \begin{axis}[
      width=\columnwidth,
      height=0.6\columnwidth,
      xlabel={\# Interpolation Points},
      ylabel={Time (ms)},
      xmode=log,
      grid=both, grid style={dashed,gray!30},
      legend style={at={(0.02,0.98)},anchor=north west,font=\footnotesize},
      cycle list name=exotic, mark options={solid}
    ]
      \addplot table[x=points,y=cpu_cubic,col sep=comma] {data/grid_A_runtimes_averaged.csv};
        \addlegendentry{CPU}
      \addplot table[x=points,y=gpu_cubic,col sep=comma] {data/grid_A_runtimes_averaged.csv};
        \addlegendentry{GPU}
    \end{axis}
  \end{tikzpicture}
  \caption{Semi-log plot of cubic interpolation for Grid A testing}
  \label{fig:cubic_perf}
\end{figure}

\begin{figure}[!t]
  \centering
  \begin{tikzpicture}
    \begin{axis}[
      width=\columnwidth,
      height=0.6\columnwidth,
      xlabel={\# Interpolation Points},
      ylabel={Time (ms)},
      xmode=log,
      grid=both, grid style={dashed,gray!30},
      legend style={at={(0.02,0.98)},anchor=north west,font=\footnotesize},
      cycle list name=exotic, mark options={solid}
    ]
      \addplot table[x=points,y=cpu_krig,col sep=comma] {data/grid_A_runtimes_averaged.csv};
        \addlegendentry{CPU}
      \addplot table[x=points,y=gpu_krig,col sep=comma] {data/grid_A_runtimes_averaged.csv};
        \addlegendentry{GPU}
    \end{axis}
  \end{tikzpicture}
  \caption{Semi-log plot of kriging interpolation for Grid A testing}
  \label{fig:kriging_perf}
\end{figure}
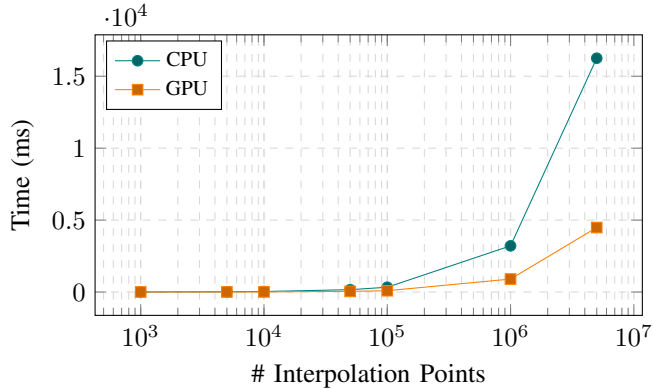

Figures \ref{fig:bilinear_perf}, \ref{fig:cubic_perf}, and \ref{fig:kriging_perf} show the runtime results of Grid A testing. For smaller batches of randomly generated interpolation points, the performance of CPU and GPU code tracks well across the three interpolation methods. However, as the number of interpolation points exceeds 100,000, the graphs diverge more clearly and the runtime for the CPU methods increases for the cubic spline and ordinary kriging methods. Note that the plots are semi-log to show exponential growth more clearly.

%— Bilinear Plot —%

%— Cubic Plot —%

%— Kriging Plot —%

One notable observation is the minimal difference in the performance of the bilinear interpolation method on the CPU and GPU across batch sizes. This is likely due to the schemes being implemented in almost the same way on both the CPU and the GPU. The similarity in their algorithm design and the simplicity of the bilinear interpolation method also explain this. However, with a more capable device that has shared memory, and by taking advantage of warp sizes, the GPU methods can attain much better performance not just on the bilinear method, but also the cubic spline and ordinary kriging methods.

\subsection{Grid B Testing}

The Grid B tests had some anomalies in their visual bathymetry graphs produced before and after interpolation. Their runtime analyses also had some interesting results. Since the testing criteria for Grid B involved real data that was collected, the interpolation methods are also evaluated on how far the calculated value was from the actual recorded value. In order to go through each different test case, the results are separated into further subsections below.

\vspace{0.2cm}

\subsubsection{Mid-Atlantic Ridge}

Table \ref{tab:mid_atlantic_errors} shows that the ordinary kriging algorithm surprisingly had the best GPU performance with CPU performance remaining pretty low for the three interpolation methods. With data obtained through a removal fraction of 0.01, this is likely due to faster searching of valid candidate neighbors and a lower quantity of fallback calculations compared to the cubic spline and bilinear methods. 

\begin{table}[!h]
  \centering
  \caption{Mid-Atlantic Ridge Accuracy \& Performance}
  \label{tab:mid_atlantic_errors}
  % scale the tabular to exactly \columnwidth, height auto (the `!`)
  \resizebox{\columnwidth}{!}{%
    \begin{tabular}{lcccccc}
      \toprule
      Metric 
        & \multicolumn{2}{c}{Bilinear} 
        & \multicolumn{2}{c}{Cubic} 
        & \multicolumn{2}{c}{Kriging} \\
      \cmidrule(lr){2-3} \cmidrule(lr){4-5} \cmidrule(lr){6-7}
         & CPU & GPU & CPU & GPU & CPU & GPU \\
      \midrule
      Time (ms) &    0.0  &   13.0   &   3.5   &   38.0  &    3.5   &    6.0   \\
      MAE       &   16.8584 & 16.8584 & 15.7884 & 15.7884 & 13.4503 & 13.4503 \\
      RMSE      &   26.5357 & 26.5357 & 27.2985 & 27.2985 & 22.6796 & 22.6796 \\
      \bottomrule
    \end{tabular}%
  }
\end{table}

However, for this region, the CPU and GPU kriging algorithms performed with much better accuracy as their error metrics were lower than the other three interpolation methods. For smaller datasets such as this one, a difference of 3-4 meters in error is not much, but can be amplified for larger datasets, or regions with greater variability.

\vspace{0.2cm}

\subsubsection{East-Pacific Rise}

Table \ref{tab:east_pacific_errors} shows the results for a region that varies more than the previous mid-Atlantic region. Keeping a removal fraction of 0.01, GPU and CPU runtime for cubic spline and kriging interpolation remain high, but the GPU kriging runtime is almost half that of the CPU runtime. Looking at the accuracy metrics, kriging clearly outperforms bilinear and cubic interpolation.

\begin{table}[!h]
  \centering
  \caption{East-Pacific Rise Accuracy \& Performance}
  \label{tab:east_pacific_errors}
  % scale the tabular to exactly \columnwidth, height auto (the `!`)
  \resizebox{\columnwidth}{!}{%
    \begin{tabular}{lcccccc}
      \toprule
      Metric 
        & \multicolumn{2}{c}{Bilinear} 
        & \multicolumn{2}{c}{Cubic} 
        & \multicolumn{2}{c}{Kriging} \\
      \cmidrule(lr){2-3} \cmidrule(lr){4-5} \cmidrule(lr){6-7}
         & CPU & GPU & CPU & GPU & CPU & GPU \\
      \midrule
      Time (ms) &   2.5 & 14.5 & 43.5 & 52.0 & 43.5 & 23.5   \\
      MAE       &   14.4228 & 14.4228 & 14.4862 & 14.4862 & 11.1512 & 11.1512 \\
      RMSE      &   24.8556 & 24.8556 & 25.9785 & 25.9785 & 19.4132 & 19.4132 \\
      \bottomrule
    \end{tabular}%
  }
\end{table}

\vspace{0.2cm}

\subsubsection{Mariana Trench}

Table \ref{tab:mariana_errors} shows the results, with removal fraction = 0.05, of interpolation in the deepest point in the ocean. In terms of interpolation time, the Mariana example follows similar trends from the previous regions. GPU kriging continues to surpass CPU kriging and GPU cubic spline interpolation. Kriging in general continues to outperform cubic spline and bilinear interpolation in accuracy of elevation measurements.

\begin{table}[!h]
  \centering
  \caption{Mariana Trench Rise Accuracy \& Performance}
  \label{tab:mariana_errors}
  % scale the tabular to exactly \columnwidth, height auto (the `!`)
  \resizebox{\columnwidth}{!}{%
    \begin{tabular}{lcccccc}
      \toprule
      Metric 
        & \multicolumn{2}{c}{Bilinear} 
        & \multicolumn{2}{c}{Cubic} 
        & \multicolumn{2}{c}{Kriging} \\
      \cmidrule(lr){2-3} \cmidrule(lr){4-5} \cmidrule(lr){6-7}
         & CPU & GPU & CPU & GPU & CPU & GPU \\
      \midrule
      Time (ms) &   9.0 & 18.5 & 127.0 & 80.5 & 141.5 & 72.5   \\
      MAE       &   33.5738 & 33.5738 & 35.7862 & 35.7862 & 28.0409 & 28.0409 \\
      RMSE      &   50.5098 & 50.5098 & 57.7545 & 57.7545 & 42.8661 & 42.8661 \\
      \bottomrule
    \end{tabular}%
  }
\end{table}

\vspace{0.2cm}

\subsubsection{Kerguelen Plateau}

The last region considered in Grid B testing, this plateau really evaluates the abilities of the interpolation methods due to its location on a tectonic plate and volcanic activity. Table \ref{tab:kerguelen_errors} shows that with a removal fraction of 0.05, the results, once again, follow a similar trend but the runtime increase more linearly compared to the other regions. For accuracy, kriging still outperforms cubic spline and bilinear interpolation, although with a smaller difference. 

\begin{table}[!h]
  \centering
  \caption{Kerguelen Plateau Accuracy \& Performance}
  \label{tab:kerguelen_errors}
  % scale the tabular to exactly \columnwidth, height auto (the `!`)
  \resizebox{\columnwidth}{!}{%
    \begin{tabular}{lcccccc}
      \toprule
      Metric 
        & \multicolumn{2}{c}{Bilinear} 
        & \multicolumn{2}{c}{Cubic} 
        & \multicolumn{2}{c}{Kriging} \\
      \cmidrule(lr){2-3} \cmidrule(lr){4-5} \cmidrule(lr){6-7}
         & CPU & GPU & CPU & GPU & CPU & GPU \\
      \midrule
      Time (ms) &   46.0 & 71.5 & 653.0 & 246.0 & 705.5 & 358.5   \\
      MAE       &   6.74368 & 6.74368 & 6.30093 & 6.30093 & 4.98024 & 4.98024 \\
      RMSE      &   15.9349 & 15.9349 & 15.0715 & 15.0715 & 11.2783 & 11.2783 \\
      \bottomrule
    \end{tabular}%
  }
\end{table}

For the Kerguelen Plateau, and the other regions tested above, it may make intuitive sense to use bilinear interpolation if the accuracy discrepancy is a tradeoff willing to be made for faster runtime. However, as described in Section III. D, bilinear interpolation has no fallback and could return erroneous elevations that are safeguarded against in the cubic spline and kriging methods. The Kerguelen tests are a good example of this because upon interpolation, there were some points that both CPU and GPU bilinear interpolation were unable to compute, resulting in holes. 

\begin{figure}[!h]
  \centering
    \includegraphics[width=\linewidth]{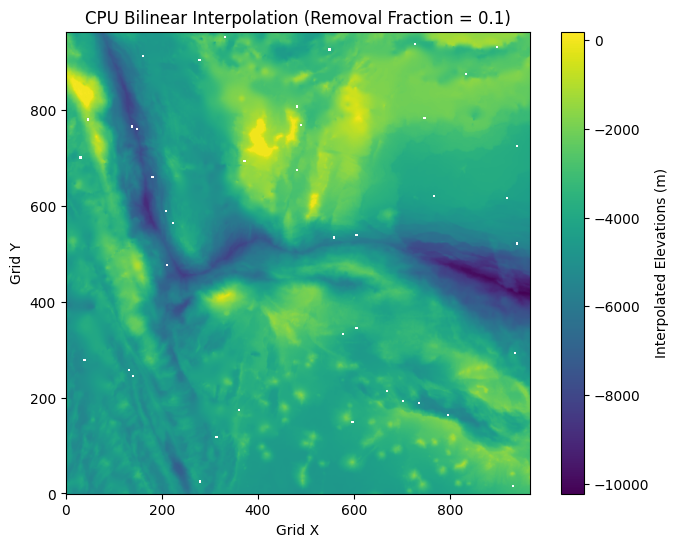}
    
    \includegraphics[width=\linewidth]{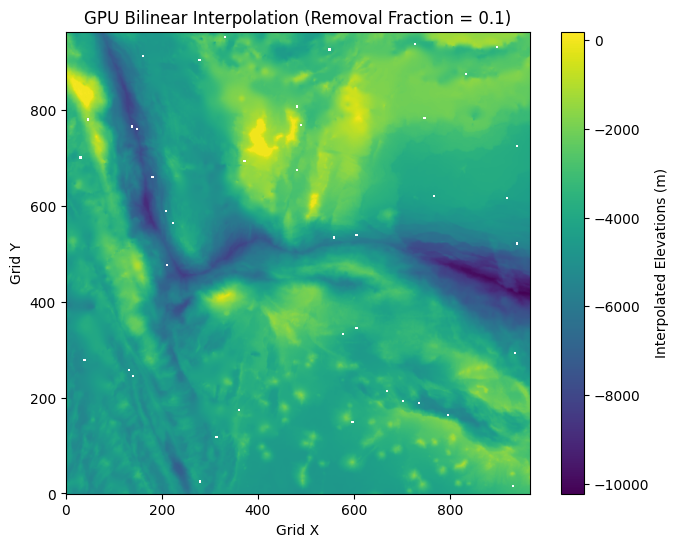}
  \caption{CPU (top) and GPU (bottom) bilinear elevation maps for the Kerguelen Plateau grid.}
  \label{fig:kerguelen}
\end{figure}

When tests for the same region were run with a higher removal fraction, such as 0.1, the quantity of holes nearly doubled, shown in Fig. \ref{fig:kerguelen}. This is a significant finding because in most use cases, the ability of the interpolation algorithms to manufacture data is of the highest importance and can not be a tradeoff. The following conclusion section discusses these tradeoffs and holistic results of the testing.

\section{Conclusion}

As is the case with most studies on the advantage of GPU computing on mathematical algorithms, testing results are judged based on tradeoffs that are willing to be made, and what factor is most important for real world applications. The following conclusion considers the application of an AUV's ability to accurately process massive amounts of data in real time and speed up the calculations from the current methods relying primarily on CPU computation.

Grid A testing found that as the batch of data increases, GPUs outperform CPUs. This is quite an intuitive finding and only confirms the preconceived notion that GPUs are built to process massive amounts of data and run similar calculations in parallel. A similar trend is seen in Grid B testing on the four regions, although the trends are less uniform and are not as clear of a sign of exponential growth as the Grid A testing examples.

However, one factor that is unanimous throughout the four regions in Grid B testing is the accuracy advantage of kriging compared to cubic spline and bilinear interpolation. This too makes intuitive sense as kriging is the more mathematically advanced algorithm of the three. Additionally, as mentioned in the Kerguelen Plateau testing section, bilinear interpolation returns faulty values when the quantity of missing data exceeds fractions as low as 0.05. One way to get around this would be to implement some other safeguard into the bilinear algorithm, but that would incur a loss in the runtime advantage the method had over cubic spline and kriging. Moreover, the holes resulting from bilinear interpolation in Grid B testing were not included in the error metrics, so the accuracy would not increase.

Summing up these findings, it is clear that even low-end, easily available GPUs like the MX550 used in this study can make a vast difference in interpolation calculation time. Paired with the accuracy of the ordinary kriging method, many of the problems that AUVs currently face with data collection, processing, and transferring can be mitigated through real-time interpolation on the AUV itself. As mentioned in the introduction sections, companies have been able to put GPU hardware on the AUVs. This allows for paralellized GPU interpolation algorithms to make a real, observable difference in the world of deep sea mapping.

\begingroup
  %----------------------------
  % 1) Hanging indent of 1.5em after the first line
  \setlength{\hangindent}{1.5em}%
  \setlength{\hangafter}{1}%
  %
  % 2) Allow “sloppy” line breaks (will avoid overruns)
  \sloppy

\endgroup

\end{document}